\documentclass[conference]{IEEEtran}
\IEEEoverridecommandlockouts
\usepackage{cite}
\usepackage{amsmath,amssymb,amsfonts}
\usepackage{algorithmic}
\usepackage{graphicx}
\usepackage{textcomp}
\usepackage{xcolor}
\def\BibTeX{{\rm B\kern-.05em{\sc i\kern-.025em b}\kern-.08em
    T\kern-.1667em\lower.7ex\hbox{E}\kern-.125emX}}
\begin{document}

\title{Embedding Culture and Grit in the Technology Acceptance Model (TAM) for Higher Education}

\author{\IEEEauthorblockN{Parvathy Panicker}
\IEEEauthorblockA{\textit{Department of Computing} \\
\textit{Goldsmiths University of London}\\
London, United Kingdom \\
ppani001@gold.ac.uk}
ORCID: 0000-0002-6934-6013
}

\maketitle

\begin{abstract}
The implementors of learning technologies within education environments often follow strategies that assume the educational environment within which they are being introduced is culturally neutral. A comprehensive literature review including 150 papers on educational technology challenges was undertaken. 

The purpose of this review is explore different  contextual challenges to the adoption of educational technology in the higher education sector. The cultural factors that define the key stakeholders (e.g., teachers, lectures, students and support staff) are often ignored when the implementation processes are undertaken. Furthermore, it is often assumed that the personnel responsible for the implementation are also culturally neutral and do not possess any attributes unique to their culture. It has been shown that cultural factors may significantly influence the implementation of learning technologies and to design strategies that fail to consider factors may  limit their efficiency and effectiveness. The challenges are interrelated and based on the findings, this review proposes a conceptual framework by integrating culture and grit into the Technology Acceptance Model(TAM) for  implementing educational technology in higher education. The framework will be useful to guide both practice and research

\end{abstract}
\section{Introduction}
Educational technology involves the integration of technologies and media in instructional contexts and processes to enhance the effectiveness of teaching-learning. A number of factors hinder implementation of educational technologies in developing countries. These factors focus on the following problems  : limited internet and bandwidth, erratic electricity supply \cite{oye2011challenges}, cost of maintenance, difficult access to technology in rural areas \cite{gunawardana2005empirical} and English proficiency \cite{qureshi2012challenges}. In addition to these practical challenges,  culture has been recognised as one of the conceptual barriers for implementing e-learning \cite{gronlund2010mobile}\cite{shraim2010learning}. Educational technologies have a great potential to support stakeholders such as students, lecturers, teachers and support staff in education. However, its effectiveness ultimately depends on the design and implementation of systems that are culturally adaptive to its stakeholders. \\
The Technology Acceptance Model(TAM) (Davis,1989) serves as a framework for technology acceptance and a number of studies have successfully extended its application to the context of electronic learning through web-based systems \cite{arenas2011cross},\cite{ turel2011integrating},\cite{ agudo2014behavioral}, \cite{tarhini2014measuring}, \cite{ros2015use}. Until now most studies have analysed design and implementation of electronic learning within the context of developed countries \cite{al2016empirical},\cite{ rutherford2017teacher}. However, there are unique pedagogical cultural challenges \cite{BhuasiriWannasiri2012CSFf}, faced by developing countries while implementing technology. Most developing countries have traditional, structured and hierarchical face-to-face classroom teaching that values rote learning, and accepting the concept of a virtual teacher delivering instructions can be a challenge. Designing systems that are culturally adaptive to the needs of indigenous cultural requirements to reduce the contextual challenge would be the direction forward. Technology acceptance Model(TAM), has been widely researched especially for the acceptance of e-learning in different cultural backgrounds. Cultural dimensions of Power Distance(PD), Uncertainty Avoidance(UA) \cite{HofstedeGeertH2010Caos} and its moderating effects on Perceived Usefulness (PU)of TAM in Lebanese culture \cite{tarhini2017examining} has been studied. Similarly, a study in Indonesia, exploring culture and Management Information System (MIS) found that long-term orientation, PD and Individualism(I) all had an effect on TAM for MIS acceptance \cite{sriwindono2012toward}. Individualistic (I) measure and UA were found to be  heavily influencing educator\textquotesingle s intention to use web-based learning tools \cite{sanchez2009exploring}. Recent studies have shown that intrinsic motivation also increases e-learning readiness. Amongst intrinsic motivation, \lq grit \rq  is defined as the \lq passion and perseverance for long-term goals\rq \cite{DuckworthAngela2016G:tp} and is associated not only with students but also with educators and lecturers. Grittier individuals have better learning satisfaction \cite{AparicioManuela2017Gitp}. The association between grit and culture is unique and this review aims to develop an extended TAM model which can form as an integral part to implementing educational technologies. \\

The objective of this review was to explore the relationship different contextual factors that influence the adoption of learning technologies. There were two research questions:(1) What are the identifiable contextual challenges when using learning technologies? ,and (2) What factors should be considered to personalise culture to motivate stakeholders to implement learning technologies in developing countries?\\

The paper is structured as follows:
Section II explains the methods used for the survey followed by a brief discussion of the contextual factors in section III. Section IV proposes a  conceptual framework of environmental challenges identified followed by section V where by a  new implementation model is proposed and conclusion is provided in Section VI. \\

\section{Methods}

A search for peer-reviewed published literature was performed using the following key words: technology and culture, electronic learning and developing countries, e-learning and developing countries, internet penetration and economy, political factors and educational technology, social factors and educational technology, economic factors and educational technology, learning technologies and political factors, electronic learning and culture learning technologies and culture, social media and educational technologies, mass media and educational technologies.
 The following databases were searched : ProQuest, Sage journals, Sage publications, Science Direct, ERIC, JSTOR and Directory of Open Access Journals(DOAJ). Google, Google Scholar, and reference lists from key studies were also used to find information relevant to this topic. One hundred and fifty  publications were reviewed, and 63 were found to be relevant to the topic. 
    \\

\section{Contextual factors in Educational TEchnology}
This section will review the different challenges the papers have addressed and categorise them accordingly.  Contextual challenges identified were- \textit{culture, societal influence, political and economic factors}.\\

\subsection{Culture}
The most commonly stated challenge concerning learning technologies with regard to context is culture. Concerns have been raised when technologies are considered to be culturally neutral and then subsequently  implemented in a culture that is different from the place of origin. 

The first issue identified here is the \textit{historical factors} contributing to the use of technology. Developing countries  have  unique cultures and tend to borrow western technology in an attempt to meet their own technological requirements. 
Developing countries follow the oral culture of rote learning, repetition and certainty through face-to-face interaction as a method of learning in primary and secondary education. Most countries that are now classed as \lq developing\rq were either under colonial rule or had several civil upsurges historically and the tradition of suppression by different political powers over time reduced the indigenous population\textquotesingle s ability to react to injustice. This suppression converted  the population into recipients of technological innovation rather than promoting invention and creativity. Colonialism foster passivity.  A system is required that acknowledges the existing educational history, thereby showing awareness that computer-based learning technologies are different from traditional class-room based teaching. The choice of \textit{language} is often found to have significant effects on learning. Learning technologies designed and implemented in western culture are predominantly in the English language, which may be different from the local cultural language. It can mostly be observed that languages introduced due to historical reasons are intended to provide functional benefits and the need to adopt/accept the nuances of new languages are lacking.  For example, during British rule in  India, English was introduced for local people to read scientific literature, that were originally published in Western countries. Any learning technology designed in a foreign language is likely to refer to what is relevant, up to date and in line with that culture. 

The concept of \textit{cultural time} as a determinant is an important factor.  The notion that \lq time is money\rq \cite{duncheon2013changing} and clock time re-enforces Western outlook over other indigenous concepts of time. The concept of cultural time is appropriate for learning technologies and differences between monochromatic and polychromatic time should be highlighted. The teaching and learning activities(TLAs) used in technological design should reflect the cultural view of time and this invariably affects learning technologies. \\

Another aspect is the co-existing relationship between culture and religion. Many societies consider \textit{religion and traditions} as integral to  their culture and the use of technology differs from one culture to another.  It can be argued that there is a need for the content to reflect religious traditions, use symbols and mottos and adopting beliefs which have more relevance for a lifestyle which closely matches their background.  \textit{Body language hermeneutics}, is relevant as body language  exhibited by all cultures differs from one another. Pictures, images, videos and emojis should be appropriate for the local convention in order not to be offensive or confused.\\

Educational background of a place should also be considered. PD experienced by students and teachers in some cultures is greater and reflects the \textit{educational culture}.  Due to PD, the teacher is considered as a single source of knowledge inside the learning environment. The knowledge of a teacher is unchallenged and it is considered disrespectful if questioned. When teachers are challenged on topics outside their academic area,  the response may be inaccurate, ignored or forgotten. Another aspect is the concept of \textit{voice}. Voice is closely related to  behaviour and beliefs and is influenced by national culture. This voice is a cultural aspect and not being heard may be experienced as a violation to certain cultural norms\cite{VanDenBosKees2010TPoV}. Finally, the \textit{institutional attitude}  towards using learning technology resources and \textit{parental concerns} around being taught using a computer rather than a teacher, weighed against expenses incurred to cover tuition costs\\
. 
\subsection{Societal influence}

Several factors that contribute towards web-based learning and its implementation indicate that there exists a clear inter-dependence between technology and society and therefore the manner in which technology is used depends upon how it is rendered. \textit{Trust} is a factor that is frequently discussed in studies about the effects on society to adapt a technology. Trust dynamics formed can be humanistic (social trust) or non-humanistic(technology trust). The relationship between trust and learning technology is rarely elaborated; the reasons for success or failure in studies are simply referred to as \lq trust-issues \rq. Another factor is \textit{personality}, which places an individual in certain dispositions under the realm of societal rules.  The relationship between cultural age, the perception about being mature, and personality determines if the individual is more or less likely to be distracted when using learning technologies.
Most often stakeholders work in institutions where  \textit{managerial support} plays an important role in adapting to new technologies; managerial support is classed as a social factor and  institutions that provide services to society assume that educational technology as a field has a social construct to cause no harm to social systems or individuals. The stakeholders of educational technologies must be provided with adequate managerial support, which is an important predictor of  the use of technology by lecturer\textquotesingle s  and those who use more technology are generally more successful in using technology in future.

Lecturers express less anxiety when more resources are available to meet  student demands. A fourth concern is the concern of \textit{job insecurity}.
Having a job in education is  not enough and  lecturers  must continually be trained in new technologies to feel confident and secure in their respective  position. Rather paradoxically, an extreme interpretation is that teachers may become  obsolete and replaced by technology. Finally, \textit{social structure} is a factor that is reported to make a difference when considering contextual factors in learning technologies. Social structure is the manner in which employees in an organisation are ranked according to hierarchy and notably, individuals with higher authority demonstrate more confidence using technology compared with lower authority, members which suggesting that more technology resources are available with the former. 

\subsection{Political and Economic factors}
Public affairs and  governmental matters at a local or national level can be influence educational technology implementation. The type of media  discussed are - social media, mass media and satellite communication; subsidies, rules and regulations, incentives and support to particular technology or a social program that favour technology in education plays an important role in its acceptance and rejection. \textit{Government-media} can influence ICT adoption and implementation. 
The relationship between different media  and its promotion by governments are rarely elaborated; the reasons for success or failure in studies are simply referred to as \lq political reasons\rq or \lq government policy\rq.  \textit{Techno-political} liberty is determined by the level of  increasing dependence of receiving nations on the technology provided by developed nations. Freedom to adapt to borrowed technology with suitable changes to reflect political differences is an important predictor of students\textquotesingle learning and those technologies that are adapted to in local emancipatory culture are likely to have a successful design and implementation strategy. A third concern is  \textit{technology diffusion} and the digital divide between countries. The rate at which learning technology is used depends on a large extent to the urban population, population age, internet costs, ICT infrastructure and market type.\\
A further aspect directly related to political factors which  impacts  adoption of learning technologies is  \textit{tertiary education}, institutions that produces high levels of quality labour-force.  Economic development of a country hugely depends on the amount of high-quality labour-force produced by the colleges and universities. High-tech field of studies are important indicators of quality labour-force.   Another important factor in the economic development is \textit{governmental programs} which include subsidies and incentives for introducing ICT and learning technologies in secondary education.  The development and implementation initiatives such as ICT4D, ICT education for girls, learning programs under direct control of government will promote a greater  number of students transfer  from secondary education to tertiary to become more  skilled and adaptable to learning based on ICT. Finally, \textit{physical infrastructure} and \textit{population density} of number of students assigned to a single or shared PC in developing countries should also be considered. \\

\section{A Conceptual implementation framework}
The previous section has shown that the three major strands of conceptual challenges (culture, social, politico- economic) are cavernous with each individually representing a set of sub challenges. 



This review has yielded nineteen challenges belonging to three main categories; challenges pertaining to culture(the customs and tradition of education system existing); social challenges(interactions exhibited when in partnered with other members of same society or institution); and politico-economic  challenges(governmental policies ). The table summarises these findings. 

\begin{enumerate}
	\item Cultural challenges
    \begin{enumerate} 
    	\item Historical factors contributing to the use of technology
        \item Language
        \item Cultural time
        \item Religion and traditions
        \item Body language hermeneutics
        \item Educational culture
        \item Institutional attitude
        \item Parental concerns
    \end{enumerate}
    \item Social challenges
        \begin{enumerate}
    	\item Trust
        \item Personality
        \item Managerial support
        \item Job insecurity
        \item Social structure
    \end{enumerate}
    \item Political and Economic challenges
     \begin{enumerate} 
    	\item Government-media type
        \item Techno-political liberty
        \item Technology diffusion
        \item Tertiary education
        \item Governmental programs
        \item Physical infrastructure and population density
    \end{enumerate}
\end{enumerate}

\section{Proposed research model}

Technology acceptance model(TAM)\cite{davis1989user} is one of the most researched models in Information Systems(IS), used to test user acceptance of technology systems. TAM is based on Theory of Reasoned Action(TRA) \cite{fishbein1975intention}. According to TAM, the main constructs are Perceived Usefulness (PU), Perceived Ease of Use(PEOU), Behavioural Intention(BI) and Actual Usage (AU). PU is defined as the subjective users\textquoteright probability that using the computer system will yield to be beneficial for promotion of one self or to climb the organisational ladder whereas PEOU  is prospective user\textquotesingle s expectation that the system is relatively easy to use. Behavioural Intention (BI) is considered to be the immediate antecedent to the actual usage of the system (AU). TAM has undergone changes in its design over the years and one such is the inclusion of the construct called Subjective Norm(SN) in TAM2 \cite{venkatesh2000theoretical}. Subjective Norm(SN) which is defined as the close circle of friends and family whose opinions the user may value and hence helps to shape one\textquotesingle s own decisions. An extended TAM  including national cultural differences was incorporated in Unified Theory of Acceptance and Use of Technology(UTAUT) \cite{venkatesh2003user} that determined the BI  of using a computer system. TAM has been much researched in mobile banking, consumer acceptance of information technology, health care and e-government services and also in evaluating students\textquoteright acceptance of tablets, electronic learning, online learning and web CT.  \\
Educational technologies are not culture neutral and there exists significant cultural measures in Higher Education sector(HE)\cite{arXiv:2005.11020}.  Hofstede\textquotesingle s  cultural measures\cite{HofstedeGeertH2010Caos} such as Power Distance (PD), Uncertainty Avoidance (UA), Individualism/Collectivism (I/C), Masculine/ Feminine(M/F) were found to have considerable effects among teachers. PD is defined as the inequality that exists in a social environment such as an institution or an organisation in which there exists a differentiation of power among individuals such that an inequality may be perceived as less or more powerful. UA is defined as shunning of ambiguity in order to look for relationships, events which makes it interpretable, predictable and follows definiteness. Individualism (I)pertains to societies in which the ties between individuals are loose: \lq everyone is expected to look after him- or herself and his or her immediate family\rq. Collectivism(C) as its opposite pertains to societies in which people from birth onward are integrated into strong, cohesive in-groups, which throughout people\textquotesingle s lifetime continue to protect them in exchange for unquestioning loyalty. A society is called Masculine(M) when emotional gender roles are clearly distinct: \lq men are supposed to be assertive, tough and focussed on material success, whereas women are supposed to be more modest, tender, and concerned with the quality of life\rq ; Feminine(F) society as \lq a society when emotional gender roles overlap: both men and women are supposed to be modest, tender and concerned with the quality of life\rq. Additionally, non-cognitive factor grit \cite{DuckworthAngela2016G:tp} which is the passion and perseverance to complete long term goals were found to have effects on the cohort\cite{arXiv:2005.11020}.  Hence the proposed CG-TAM model considers both cultural and intrinsic motivational factors as its constructs.

 \section{ CG-TAM model}
This paper proposes a research model for learning technologies drawing from literature and 
Different contextual factors have been studied in a cohort of lecturers in the Indian higher education sector which suggested that culture and grit have significant impact while implementing technology in teaching.
Results from this  study suggest that these contextual factors need to be considered for successful acceptance of educational technology. It is therefore proposed that these contextual factors are integrated
into the existing TAM model to include these additional dimensions. The proposed CG-TAM model includes  cultural, contextual and grit factors as moderators.\\

\begin{table}
\begin{tabular}{ll}	\hline
\emph{Construct} & \emph{Concept} \\ \hline 
Power Distance (PD)&\\
Uncertainty Avoidance(UA)&Hofstede\\
Individualism/Collectivism(I/C)&Hofstede\\
Masculine/Feminine(M/F)&Hofstede\\
Perceived Usefulness(PU)&TAM\\
Perceived Ease of USe(PEOU)&TAM\\
Subjective Norm(SN)&TAM\\
Behavioural Intention(BI)&UTAUT\\
Actual Use(AU)&UTAUT\\
Grit&Duckworth\\
\end{tabular}
\caption{\label{tab:table-name}Different  Constructs have been used in designing CG-TAM model}
\end{table}

A general assumption would be to start out assuming that users with high PD would be likely to be influenced by views of lecturers and teachers whom the user perceive as superior to them. \cite{srite2006role} measured culture and in one part of their study  found that relationship between SN and BI was stronger for individuals with low PD cultural value. To the contrary, \cite{tarhini2017examining} found the relationship between SN and BI to be stronger for high PD in Lebanese culture. \cite{sadeghi2014impact} defined PD to have negative effects on PU and PEOU for high PD Iranian culture where SN was influenced by traditional characteristics of dominating power. \\

Similarly, an increase in UA has shown to be associated with a greater adoption of open source software  \cite{qu2011multi}. \cite{sanchez2009exploring} measured  UA in the two European cultures- Nordic and Mediterranean- to understand the determining factors that influenced the intention to use web based electronic learning, and concluded that educators from moderate to low UA perceived ICT use as intrinsically motivating with little need for social approval and acceptance of innovative ideas. \cite{tarhini2017examining} demonstrated that for electronic learning, stronger relationship exists between PU and BI  for students\textquotesingle  with greater UA. On the other hand, UA did not moderate the relationship between PEOU and BI. Therefore, in this theoretical model it is proposed to use UA, PD, I/C and M/F as cultural moderators. Additionally,  this review identified environmental factors to be affecting the use of technology and therefore 19 environmental factors classified under cultural, social, politico-economic categories and grit have also been included in the CG-TAM model. \\

\begin{figure}
\includegraphics[width=80mm, scale=0.5]{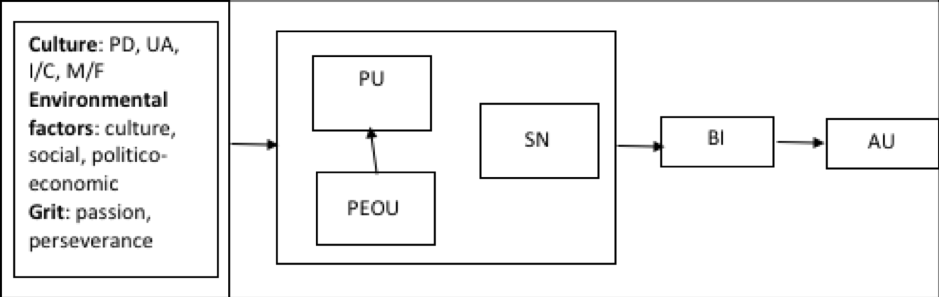}
  \caption{Proposed CG-TAM model that includes culture, grit and environment}
\end{figure}

\section{conclusion}

This paper set out to answer two research questions: \lq What are the identifiable contextual challenges when using learning technologies?\rq and \lq  What factors should be considered to personalise culture to motivate stakeholders to implement learning technologies in developing countries?\rq\\
This paper has identified nineteen contextual challenges belonging to three main categories:\\

\begin{enumerate}
	\item Cultural challenges
	\item Social challenges
	\item Politico-economic challenges
\end{enumerate}


This review has shown that while challenges exist using learning technologies in different cultures, it is crucial for stakeholders and implementors to understand the culture of the educational environment in order to make the technology efficient. Following the stages of technological advancement in different developing countries since the middle of the  20th century, the system being carried out reflects historical educational methods of teaching and the existing educational culture. 

For example, it was found that most lecturers in higher education were not opposed to the idea of lecturers being at the centre stage and directing the learning process. Some of the concerns raised were distraction and superficial knowledge about a topic. A step towards adaptations to make educational system more culture inclusive requires a  change in perception of  stakeholders and adapting to asynchronised, polychromatic use of time during classroom teaching; these changes will influence in shaping of educational technologies that are culture inclusive. \\

This review presents several implications for theory and practice.  The model adds additional dimensions of culture, contextual factors and grit in TAM.  Furthermore, these moderators are antecedent to BI in the context of the learning technology framework. The  relationship between these moderators in the context of TAM has not been explored in relation to  educational  technologies.\\

To address the challenges associated with technology acceptance when implementing and designing learning technologies in developing countries, this review propose to extend the TAM model by including moderators such as UA, PD, I/C, M/F, contextual factors and grit, and TAM model constructs such as PU, PEOU and SN; these all need to be examined in different cultural contexts. Although TAM and other user acceptance models have been validated empirically, the addition of cultural and psychological factors enhance the explanatory power of pure technology driven models and highlights a holistic approach to implementing learning technologies in developing countries. \\

\section{acknowledgement}
The author acknowledges Dr. James Ohene-Djan, Goldsmiths, University of London for input received during the design of CG-TAM model. 

\bibliography{model_ref}
\bibliographystyle{apalike}

\end{document}